\documentclass[paper]{revtex4}

\usepackage{graphicx}
\usepackage{epsfig}
\usepackage{fancyhdr}
\usepackage{axodraw}
\usepackage{empheq}
\usepackage{mathbbol}
\usepackage{wasysym}
\usepackage{pstricks}
\usepackage{color}
\usepackage{bbm}
\usepackage{bm}
\usepackage[utf8x]{inputenc}
\usepackage{amssymb,amsmath}
\usepackage{epstopdf}
\usepackage{float}
\usepackage{color}
\usepackage{subfig}
\graphicspath{{./immagini/}}
\usepackage{hyperref}

\begin{document}

\title{A Mechanism for Hadron Molecule Production in $p\bar p(p)$ Collisions}


\author{A Esposito$^{*}$, F Piccinini$^{\dag}$,  A Pilloni$^{*}$ and AD Polosa$^{*}$}
\affiliation{
$^*$Dipartimento di Fisica, Sapienza Universit\`a di Roma, Piazzale A Moro 2, Roma, I-00185, Italy\\
$^\dag$ INFN, Sezione di Pavia, Via Bassi 6, I-27100, Pavia, Italy}
\begin{abstract}
We propose a mechanism allowing the formation of loosely bound molecules of charmed mesons in
high energy proton-(anti)proton collisions.  \newline\newline
\end{abstract}
\maketitle

{\bf \emph{Introduction}}. 
The problem of understanding the loosely bound hadron molecule formation in $p\bar p(p)$ collisions at Tevatron and LHC energies is still open.
A recent measurement by the CMS Collaboration~\cite{cms} basically confirms, at higher energies, older Tevatron results on the prompt production of $X(3872)$ which were first addressed in~\cite{noiben}. 
Looking at these new results~\cite{cms},  the questions remain the same as those raised in~\cite{noiben}: how is that possible that a very long lived molecule of a $D^0$ and a $D^{*0}$ meson, with binding energy compatible with {\it zero}, could be formed within the bulk of the hadrons ejected in very high energy $p\bar p(p)$  collisions? Is it the $X(3872)$ that molecule? 

The reply given in~\cite{noiben}  to the former question  was sharply negative. In that paper we performed numerical simulations with standard hadronization algorithms (Herwig and Pythia) tuned to fit  data on the production of open charm mesons and sought $D^0\bar D^{*0}$ pairs with  reasonably low relative momentum in their centre of mass  so as to be eligible candidates for becoming molecular loosely bound states. The number of selected pairs allowed to estimate an upper bound on the  {\it prompt}~\footnote{{\it i.e.} not produced in $B$ decays but at the hadron collision vertex.} production cross section of the $X(3872)$  which was found to be at least 30 times smaller than the experimental value. 

Our analysis was reproduced, with similar results, in~\cite{abrat}, where it was also observed that a more appropriate treatment of Tevatron data would rather indicate a discrepancy with theoretical expectations by a factor of 300.

Such a gap did not seem to be unbridgeable to the authors of~\cite{abrat}, who resorted to final state interaction (FSI) mechanisms in the $D^0\bar D^{*0}$ system  in order to improve the theoretical cross section up to the experimental value. The approach there used was criticised in~\cite{noiben2} leaving the controversy somewhat unsolved~\cite{uns}. 

%A prediction of the prompt production cross section of $X$ at the LHC was also presented in~\cite{abrat} as a function of $p_\perp$. As can be read in~\cite{cms}, this prediction is off from CMS measurements  by only a factor of four, which, in our view, would be an excellent result if only we could actually rely on the approach proposed to drastically reduce the discrepancy with Tevatron data - once filled the gap with Tevatron results, computing the cross section at the LHC is only a matter of changing the energy in the hadronization Monte Carlo.  

{\bf \emph{Molecular $X(3872)$}}.
On the other hand, during the last few years, the idea of a molecular $X$, in diverse incarnations~\cite{moles}, has been corroborated by the lack of observation of its nearly degenerate charged partners, required by the antagonist tetraquark model~\cite{mainoi}.  For these reasons we come back here to the problem of the $X$ formation in high energy hadron collisions being motivated by a completely different approach.
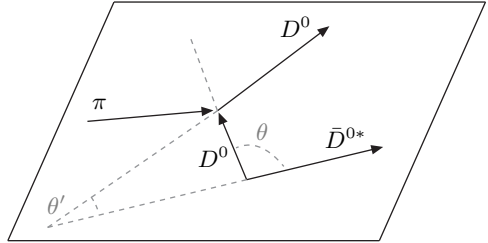
\begin{figure}[htb!]
\begin{center}
\begin{picture}(180,90)(0,0)
 \Line(0,0)(40,90)
 \Line(40,90)(180,90)
 \Line(180,90)(140,0)
 \Line(140,0)(0,0)
 \LongArrow(90,23)(140,35)%D*
 \LongArrow(90,23)(80,47)%D
 \LongArrow(30,45)(76,49)%pi
 \LongArrow(79,49)(120,80)%D'
 \Text(120,35)[bl]{$\bar D^{0*}$}
 \Text(84,28)[br]{$D^{0}$}
 \Text(32,50)[bl]{$\pi$}
 \Text(115,77)[br]{$D^{0}$}
 \SetColor{Gray}
 \DashLine(15,5)(90,23){2}%D*
 %\DashCArc(90,23)(15,20,113){2}
 \DashCurve{(86,35)(92,36)(104,28)}{2}
 \DashCArc(15,5)(20,20,35){2}
 \DashLine(79,49)(69,76){2}
 \DashLine(15,5)(79,49){2}%D*
 \Text(94,38)[bl]{\color{gray}{$\theta$}}
 \Text(15,10)[bl]{\color{gray}{$\theta^\prime$}}
\end{picture}
\end{center}
\caption{The elastic scattering of a $D^0$ (or $D^{*0}$) with a pion among those produced in hadronization could reduce the relative momentum $\bm k_0$ in the centre of mass of the $D^0\bar D^{0*}$ pair.  }
\label{angoli}
\end{figure}

In our view the $X$ could rather be the meson-molecule analogue of the stable deuterium.

Given the large number of pions produced in the neighbourhood of the open charm meson pairs in momentum phase space, it is plausible that some of  those pions could scatter  elastically on the $D^0$ or $D^{*0}$ component of the would-be-molecule  changing the relative momentum in the centre of mass of the pair, $\bm k_0$,  towards lower values - see Fig.~\ref{angoli}. We can assume the initial total energy ${\mathcal E}$ of the pair to be positive. However, if  $k_0=|\bm k_0|$ gets smaller due to an interaction with the pion, ${\mathcal E}$ might be found shifted down to some {\it negative} - close to zero - value, provided that the $D^0\bar D^{*0}$ pair is under the influence of some (unknown) attractive potential, say a square well
potential, similar to the simplest description of deuterium.
%If this happens within a  time of few fermi's after hadronization, the $D^0\bar D^{*0}$ pair could still be sensible to some interaction potential (here we think to a square well, as in the simplest description of deuterium) and, because of the lowered $\bm k_0$, the total energy ${\mathcal E}$, which is  could decrease to some {\it negative}, close to zero, value. 
%to lower values would translate into changing the kinetic energy of the pair towards lower values and possibly the total energy ${\mathcal E}$ to a {\it negative} value, close to zero, in the potential well.  

In these respects the $X$ would be a genuine, negative energy, bound state of $D^0\bar D^{*0}$ whose lifetime is entirely regulated by the lifetime of the shorter lived component $D^{*0}$; we would  estimate then  a total width  \mbox{$\Gamma_{\mathrm{tot}}(X)\simeq 65$~keV}~\cite{braatenk}.  There are no energetic arguments to stabilize the $D^*$ in the attractive potential. 

Such a mechanism is therefore somewhat opposite to the one based on FSI, where the $D^0\bar D^{*0}$ pair should rescatter remaining  isolated from other hadrons potentially produced close in phase space~\cite{abrat, noiben2}. %In our view, on the contrary, these hadrons make the loosely bound molecule possible.  

\begin{figure}[h!]
\begin{center}
\epsfig{height=4truecm, width=6truecm,figure=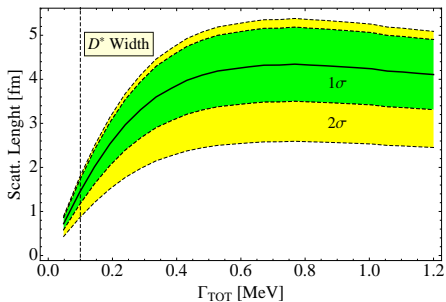}      
\caption{\small Scattering length for the $D^0\bar D^{*0}\to X\to D^0\bar D^{*0}$ process as a function of the $X$ total width. The initial pairs are selected with $|\bm k_0|<50$~MeV which, in our simulations, represent a few parts over $10^5$ of the total. 
%The maximum scattering length obtained has to be confronted with the 12~fm expected for  $D^0\bar D^{*0}\to  D^0\bar D^{*0}$ scattering with a shallow bound state. 
The error bands account for uncertainties on the data we used.}%Green and yellow bands are at 1 and 2 sigmas. }
\label{aa}
\end{center}
\end{figure}

One more reason to pursue the approach  described above is that the  resonant scattering $D^0\bar D^{*0}\to X\to D^0\bar D^{*0}$ is difficult to be reconciled with the general expectations that can be drawn  for the total scattering cross section of two particles allowing a {\it shallow bound state} with energy $|{\mathcal E}|\sim 0$,  as described in  the `Low equation' formalism, see~\cite{wei}. Resonant scattering $D^0\bar D^{*0}\to X\to D^0\bar D^{*0}$ can be computed using available data on $X$ decay branching fractions (in order to compute the $g_{_{XDD^*}}$ coupling)  and 
averaging the cross section, $\langle\sigma\rangle_f$, with the distribution $f(k_0)$ of $D^0\bar D^{*0}$ pairs  obtained by hadronization algorithms~\footnote{
$\langle\sigma\rangle_f=\int_{k_0<\Lambda} dk_0 f(k_0)\sigma(k_0)/\int_{k_0<\Lambda}  dk_0 f(k_0)$.
}.  It is only when $k_0$ is smaller than some critical value $\Lambda$ that the resonant scattering into $X$ has a non negligible probability to occur. We find a scattering legth $a$ of about $4$~fm for a total width $\Gamma(X)\simeq 1.2$~MeV (the $g_{_{XDD^*}}$ 
coupling is a function of the $X$ total width) and $\Lambda=50$~MeV. The scattering length $a$
decreases for smaller values of the total width - see Fig.~\ref{aa}. 
Shifting $\Lambda$ towards higher values, $ \Lambda\sim 10$~GeV, $a$ decreases to few cents of a fermi.

On the other hand, the scattering length expected for scattering with a shallow bound state is \mbox{$a=\hslash/\sqrt{2\mu|{\mathcal E}|}\simeq 12$~fm} (${\mathcal E}\simeq-0.14$~MeV).
%, giving a remarkably high value for the total cross section $\sigma_{\text{tot}}\simeq 16$~barn, given the very small binding energy. 
Such a result, as discussed in~\cite{wei}, is independent on the (unknown) scattering potential. 
%and has to be confronted with the maximum scattering length we find   for the process  $D^0\bar D^{*0}\to X\to D^0\bar D^{*0}$ which is about $4$~fm, as obtained for a total width $\Gamma(X)\simeq 1.2$~MeV (the $g_{_{XDD^*}}$ coupling is a function of the $X$ total width). See Fig.~\ref{aa}. 

{\bf \emph{Analysis method}}. 
The binding energy of the $X$ is estimated from the mass difference with its constituents ${\mathcal E}\simeq-0.14\pm 0.22$~MeV. A discrete level at this energy (take the central value) can be accommodated in a square well with a depth of about $-7$~MeV~\footnote{-20~MeV in the case of deuterium.} and a range $r_0\simeq 3$~fm. 

Let $\psi_0$ be the wave function associated to this level. The average size of the molecule is found to be \mbox{$\sqrt{\langle r^2\rangle_{\psi_0}}\simeq 10$~fm}  and a value of  $\sqrt{\langle \bm k_0^2\rangle_{\psi_0}}\simeq 50$~MeV is  determined. Those pions  scattering elastically on $D^0$ or $D^{*0}$ and making the $k_0$ of the pair lower than 50~MeV are  able to drop the total energy down to ${\mathcal E}$ and form a genuine $D^0\bar D^{*0}$ bound state. 
It is our purpose here to seek  such pions and to study numerically their elastic interactions with the $D^0$ or $D^{*0}$ mesons adapting standard hadronization tools such as Herwig and Pythia.  

As discussed first in~\cite{noiben}, the spectrum of $D^0\bar D^{*0}$ pairs can be represented by a monotonically rising histogram in  $k_0$. Because of the interaction with  pions, pairs with high relative COM (centre of mass) momenta, the majority, could either  be pushed to  higher momenta or to lower ones.  If even a small part of them were rearranged within lower relative momenta, there could be a significant effect of feed-down of pairs towards lower bins, even in the far low energy region below 50~MeV. Populating that region means increasing the formation probability of the loosely bound $X$. 

To perform a first qualitative exploration of this phenomenon, we start by generating samples of  $p\bar p\to  c\bar c$ events in Herwig and Pythia, at Tevatron COM energies ($\sqrt{s}=1.96$~TeV). We list the events containing $D^0 \bar D^{0*}$ (resp. $\bar D^0 D^{0*}$)
as a function of $k_0$.  The  cuts imposed  at parton level are: $p^\text{part}_\perp > 2$ GeV and $\left|y^\text{part}\right| < 6$.  

The distributions $d\sigma/d(\Delta \varphi)$, where $\Delta\varphi$ is the difference in azimuthal angles between $D^0$ and $D^{*-}$,  as discussed in~\cite{noiben}, are reproduced by choosing the following cuts on the final mesons: open charm meson pairs have $5.5~\text{GeV}<p_\perp < 20~\text{GeV}$ and $\left|y\right| < 1$. These cuts allow to reproduce very well CDF data on $d\sigma/d(\Delta \varphi)$ if a full Quantum Chromodynamics  (QCD)  generation of events is performed ($c\bar c+gg+ gq+ qq...$). 
$D^0\bar D^{*0}$ pairs in the bin $\Delta\varphi =[0^\circ,18^\circ]$ are the main would-be-molecule candidates.  We observe here that the numerical generation of  $p\bar p\to  c\bar c$ partially fills ($\approx 10\%$) the $\Delta\varphi =[0^\circ,18^\circ]$ bin with respect to the full QCD one. In addition, in the central region, which is enforced by the cuts, we have to match our results with those of some Matrix Element Monte Carlo, like Alpgen~\cite{alp}, more than just using shower algorithms. We will present the results of the full QCD simulation, which is much more time consuming, in a future paper. 

%Hadrons produced in the numerical simulation form, initially,  a `fireball' of a few fm's - the larger is the angle between two hadrons, the farther away they will be on the surface of the ball. 
To optimize the selection of events, we choose the 10 most complanar pions to the $D^0\bar D^{*0}$ plane, then we randomly choose the meson the pion will interact with (say the $D^0$), and finally we select the most parallel pion to the non-interacting meson (say the $\bar D^{*0}$) - see Fig.~\ref{angoli}. In physical events, we expect such a pion to be the most effective one to the phenomenon we are describing.

The elastic interactions with the pions are regulated in the $\pi D^0$ COM by the matrix elements
\begin{eqnarray}
&& \langle \pi(p) D(q) | D^*(P,\eta)\rangle = g_{_{\pi D D^*}}\, \eta \cdot p\notag\\
 &&\langle \pi(p) D^*(q,\lambda) | D^*(P,\eta)\rangle = \frac{g_{_{ \pi D^* D^*}}}{M_{D^*}}\, \epsilon_{\alpha\beta\gamma\delta} \lambda^{\alpha} \eta^{\beta} p^{\gamma} q^{\delta}\notag
\end{eqnarray}
where the couplings used are $g_{_{\pi D D^*}} \approx 11$, $g_{_{\pi D^* D^*}} \approx 17$, see~\cite{casal}.
After the interaction with the pion has taken place in the COM $D^0\pi$ frame, we boost back the $D^0$ in the laboratory (LAB) frame and check if the `new' $D^0\bar D^{*0}$ pair passes the cuts we fixed for the final meson pairs. 

\begin{figure}[htb!]
\begin{minipage}[t]{7truecm} % A minipage that covers half the page
\centering
\includegraphics[width=7truecm]{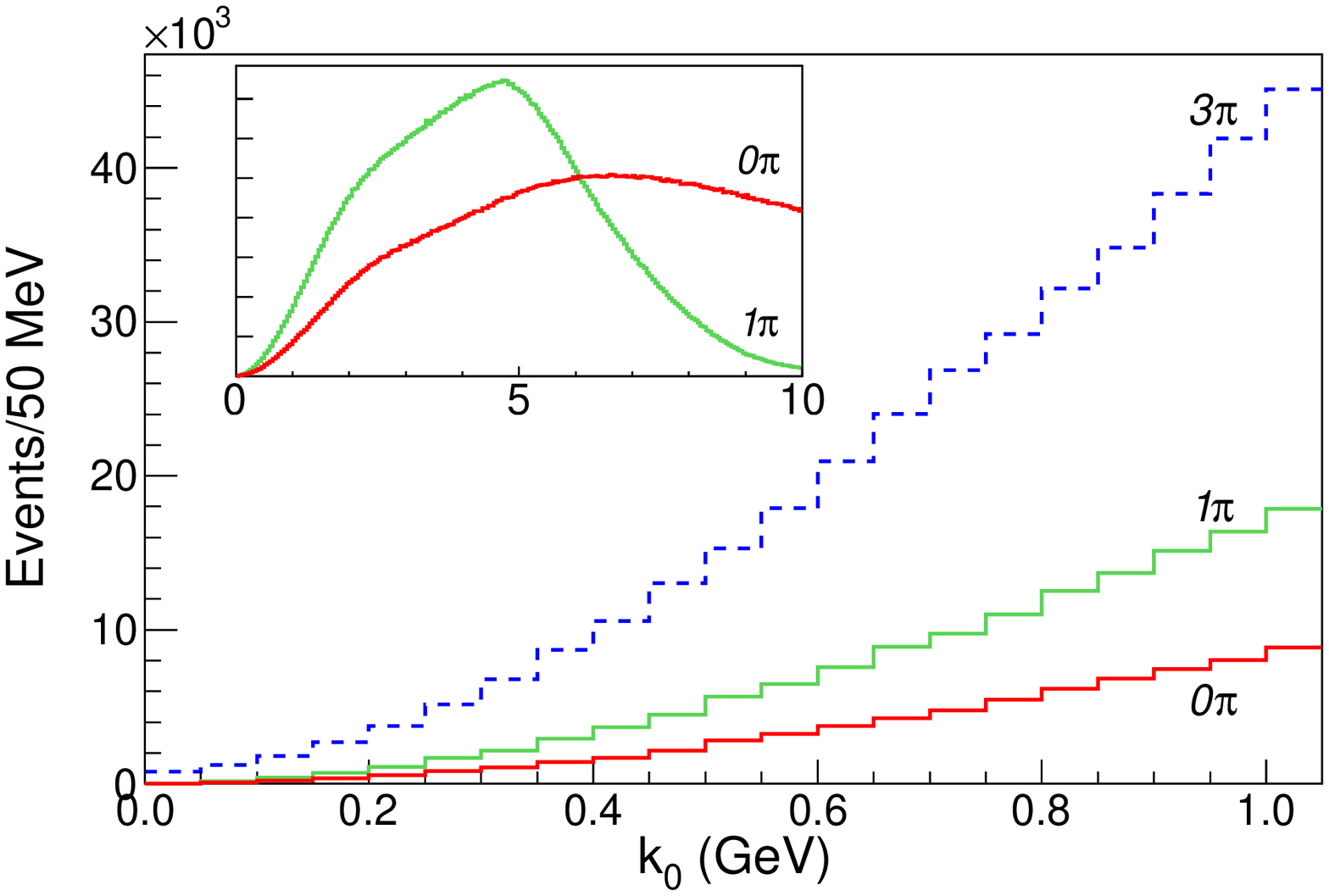}
\end{minipage}
\hspace{1truecm} % To get a little bit of space between the figures
\begin{minipage}[t]{7truecm}
\centering
\includegraphics[width=7truecm]{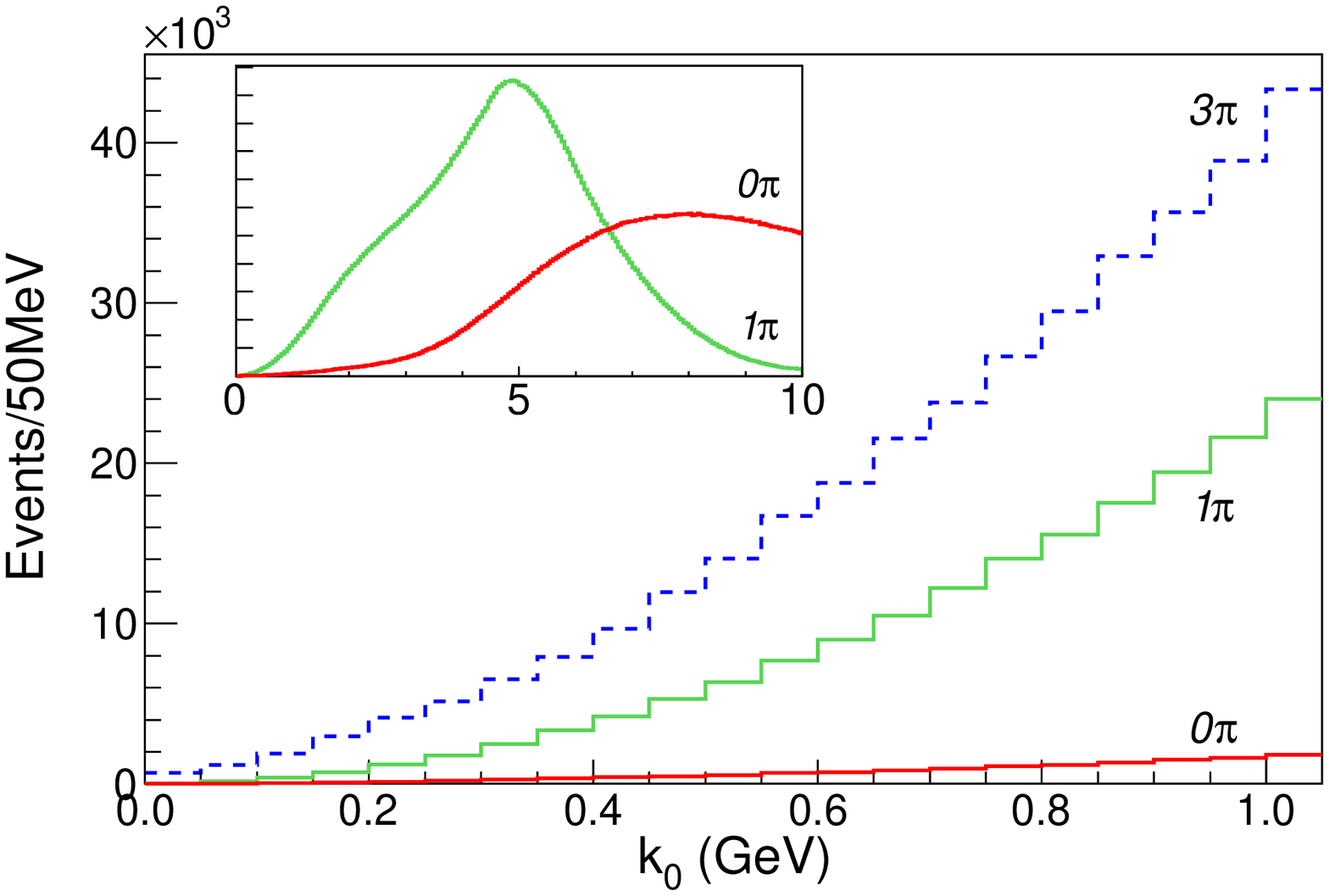}
\end{minipage}
\caption{\small Number of $D^0\bar D^{*0}$ pairs (events) counted with Herwig (left panel) and Pythia (right panel) when generating $10^{10}$ $p\bar p\to c \bar c$ events at $\sqrt{s}=1.96$~TeV with the cuts on partons and hadrons described in the text. The $0\pi$ histogram reproduces the shape found in~\cite{noiben}. The histograms named $1\pi$ and $3\pi$ are related to the elastic scattering of open charm mesons with one or three pions selected as described above. In the insects we report a broader $k_0$ range.}
\label{results}
\end{figure}

We can trace, event by event, the variation $\bm k_0\to \bm k_0^\prime$ 
of each $D^0\bar D^{*0}$ pair filling a 2D histogram  of transition probabilities $\mathcal{P}\left(k_0,\Delta  k_0\right)$.  Since the interaction with pions can change the $p_\perp$ and $y$ of the molecule, a pair might fail the strict meson cuts before the interaction and pass them after it (a `gained' would-be-molecule) and viceversa (a `lost' one): see Fig.~\ref{results}.
%If we histogram the $\bm k_0$ of `lost' would-be molecules (resp. the $\bm k^\prime_0$ of `gained' ones), we find very similar results to those obtained for  standard pairs.  This means that $p_\perp$ and $y$ are independent from $\bm k_0$. 
%Given that the distribution is just the same as the full histogram, the only additional information is the number of `lost' and `gained' molecules, we call them $l_1$ and $g_1$.
%The mechanism shows to be successful at depleting the bins in the $\bm k_0>6$~GeV region filling the lower energy ones. 

The open charm mesons might interact with pions more than once before a molecule is formed.
Roughly speaking the $\pi D^0\to \pi D^0$ scattering is proportional to $g_{{\pi D D^*}}^4$ whereas the $D^*\to D\pi$ decay is `slower' by  $g_{{\pi D D^*}}^2$~\footnote{
We might say that $\tau_{\text{scatt}}\sim1/(\rho v \sigma)\approx M^2/(g^4 (200)^3 \text{MeV}^3)\approx 1/(10^4 g^2)\;  \text{MeV}^{-1}$ where we used the $\pi D$ reduced mass for $M$. On the other hand $\tau_{\text{dec}}\sim 24\pi M_D^2/(g^2 
|\bm p^*|^{3})$ where $|\bm p^*|$ is the decay momentum. Thus $\tau_{\text{dec}}\sim 10^4/g^2$~MeV$^{-1}$. 
}. %This means that as long as pions and $D$'s are not too far apart during the fireball expansion, it is reasonable that they can keep on interacting. 
We assume that a single $D^0(D^{*0})$ might scatter, on average,  with $2\div 3$ pions before the relative distances among the flying-out hadrons are such that the interactions are suppressed~\footnote{
The hadronization time of a  $c$ or $q$ in the LAB frame is $t_{\text{had}}={\cal E}/m \; R$ where $R\approx 1$~fm and the mass is meant to be the constituent one. From our simulations we estimate that,  at the formation time of pions, the $D\bar D^{*}$ pair and the pions will be distributed on  {\it spherical segments} - around $D$ and $D^*$ - of an expanding sphere.  Using our simulations we estimate $2\div 3$ pions per~$\pi r_0^2$~fm$^2$ around $D$ or $D^*$.}.  
 
Therefore, for each pair, we wish to evaluate $ k^{(n)}_0$ after $n$ interactions. We do it according to the probability distribution functions (PDF) as extracted from 
$\mathcal{P}\left(k_0,\Delta k_0\right)$.
We build a set of PDFs $\mathcal{P}_i\left(\Delta k_0\right)$ for each bin $i$ in  $d\sigma/dk_0$. We assume that the PDFs will be the same for  all the interactions, like in a Markov chain. 
For each event we have a $k^{(n)}_0$,  falling in some particular bin $i$.  We randomly extract a $\Delta k_0$ according to the distribution $\mathcal{P}_i\left(\Delta k_0\right)$ and sum $| k^{(n)}_0 + \Delta k_0|=k^{(n+1)}_0$ thus producing a new histogram. 

We must also take into account the  `lost' and `gained' would-be-molecules. In each iteration, we generate the number of `lost' and `gained'  ones, $l^{(n)}$, $g^{(n)}$, according to  Poissonian distributions with mean values $l^{(1)}$, $g^{(1)}$. We implement the following algorithm:  
$i)$ before the $n$-th interaction, we drop out a number $l^{(n)}$ of pairs, $ii)$  we produce the new histogram as a result of the interaction with one more pion, $iii)$ after that, we decide to `gain' a $g^{(n)}$ number of pairs.

{\bf \emph{Results}}. 
The results are showed in Fig.~\ref{results}. The bin we are more interested in is the first one, with \mbox{$k_0<50$~MeV}. The number of pairs obtained for that bin are reported in Table~I.
As one can see from these plots the feed-down mechanism towards lower relative momentum bins is very effective once the interaction of a $D^0$ or a $D^{*0}$ with a pion from the hadronization is taken into account. The effect gets magnified if successive interactions are allowed (up to three). 
In the insects we show a broader range in $k_0$. It is evident here that the elastic scattering with a pion is also causing a net increase of would-be-molecule pairs: it forces a number  of pairs to pass the $p_\perp > 5$ GeV and $\left|y\right| < 0.6$ cuts, which otherwise would be failed.

\begin{table}[htb!]
\begin{center}
\begin{tabular}{ l | l | c | r }
  $k_0<50$~MeV & ~~$ 0\pi$~~ & ~~$1\pi$~~ & ~~$3\pi$~~ \\
 \hline            
\text{Herwig}&  ~~10 & 19 & 802 \\
  \text{Pythia}& ~~3 & 21 & 814 \\
  \hline  
\end{tabular}
\end{center}
\caption{The population of the the $k_0<50$~MeV bin ($D^0\bar D^{*0}$ pairs), after $0,1,3$, $\pi D^{(*)}$ interactions.}
\label{tavola}
\end{table}

The results showed in Table~\ref{tavola} are indicating qualitatively that the mechanism  described in this letter indeed occurs in numerical simulations of $p\bar p$ collisions and might play an important role in physical events. For a full determination of prompt production cross sections we need to switch from $p\bar p\to c\bar c $ to the full QCD generation $pp\to c\bar c + gg + g q+ q q ...$ which is a harder task in terms of numerical computation, yet, from the exploration here reported, we have a clear clue on what to expect.    

{\bf \emph{Conclusions}}.
We have presented a new mechanism to explain the prompt formation of loosely bound  open charm meson molecules at hadron colliders as induced by elastic scattering with comoving pions.
Simplified numerical simulations show that pions produced in hadronization  might be effective at decresing the relative momentum in the center of mass of the $D$ meson pair which, if under the influence of an attractive potential, might therefore  be found at some small negative energy, like in a shallow bound state in a potential well. 
Such a bound state will have a lifetime which is as long as the $D^{*0}$ one, $\Gamma\sim 65$~keV, still well below actual experimental resolution. 
With the results of the full numerical simulations we will provide expected prompt cross sections for the production of the $X(3872)$ at the LHC. 

Considering the known limits of the available hadronization models, the results of numerical simulations have to be taken as compelling but qualitative descriptions of the suggested mechanism. We believe that several more investigations in this direction are possible.

{\bf \emph{Acknowledgements}}. A.P. wishes to thank E. Braaten for stimulating discussion.

\end{document}